\documentclass[sigconf]{acmart}

\usepackage{booktabs} 

\usepackage[utf8]{inputenc} 
\usepackage{csquotes} 
\usepackage{enumitem} 

\let\URL\url
\makeatletter
\def\url{\begingroup \catcode`\%=12\catcode`\#=12\catcode`\?=12\relax\printurl}
\def\printurl#1{\@URL#1 \@nil\endgroup}
\def\@URL#1 #2\@nil{\URL{#1}\ifx\relax#2\relax \else; \url{#2\relax}\fi}
\makeatother

\newcommand{\codeee}[1]{%
  \begingroup
  \ttfamily
  \begingroup\lccode`~=`:\lowercase{\endgroup\def~}{:\discretionary{}{}{}}%
  \begingroup\lccode`~=`_\lowercase{\endgroup\def~}{_\discretionary{}{}{}}%
  \catcode`:=\active\catcode`_=\active
  \scantokens{#1\noexpand}%
  \endgroup
}
\newcommand{\code}[1]{\mbox{\codeee{#1}}}
\newcommand{\shrink}{\vspace{-1ex}}

\setcopyright{rightsretained}

\acmDOI{10.475/123_4}

\acmISBN{123-4567-24-567/08/06}

\acmConference[Drift-a-LOD'17]{Drift-a-LOD workshop}{September 2017}{Amsterdam, Netherlands} 
\acmYear{2017}
\copyrightyear{2017}

\acmPrice{2/3}

\begin{document}
\title{Finding Talk About the Past in the Discourse of Non-Historians}

\author{Alex Olieman}
\orcid{0000-0001-6533-5328}
\affiliation{%
  \institution{University of Amsterdam}
}
\email{olieman@uva.nl}
\affiliation{%
  \institution{Stamkracht BV}
}
\email{alex@stamkracht.com}

\author{Kaspar Beelen}
\affiliation{%
  \institution{University of Amsterdam}
}
\email{k.beelen@uva.nl}

\author{Jaap Kamps}
\orcid{0000-0002-6614-0087}
\affiliation{%
  \institution{University of Amsterdam}
}
\email{kamps@uva.nl}


\begin{abstract}
A heightened interest in the presence of the past has given rise to the new field of memory studies, but there is a lack of search and research tools to support studying how and why the past is evoked in diachronic discourses.
Searching for temporal references is not straightforward. It entails bridging the gap between conceptually-based information needs on one side, and term-based inverted indexes on the other.
Our approach enables the search for references to (intersubjective) historical periods in diachronic corpora. It consists of a semantically-enhanced search engine that is able to find references to many entities at a time, which is combined with a novel interface that invites its user to actively sculpt the search result set.
Until now we have been concerned mostly with user-friendly retrieval and selection of sources, but our tool can also contribute to existing efforts to create reusable linked data from and for research in the humanities.

\end{abstract}

%
%



\maketitle

\section{Introduction}

There has been a monumental shift from the future to the past in the cultural orientation of Western societies, starting in the 1980s. In a sense, the past has increasingly gained in presence: in the literary and artistic expressions of (traumatic) memories that cannot be contained by the evidence that forms the basis of historical studies, the proliferation of museums and archives, and the commodification of the past as marked by docudramas, historically-themed amusement parks, and memorabilia of pasts that never existed \cite{Huyssen2000}. Many of the resulting representations of segments of the past serve a deeply different purpose than the representations created by professional historians. Under the influences of positivist science and literary realism in the 19th century, history was detached from ``its former habitation in rhetoric'' \cite[p. 10]{white2014practical} and has developed a rigorous manner of dealing with sources and evidence that leads to the production of distinctly historical accounts of past events. While changes in how historians have interpreted past events has been widely studied (e.g. under the banner of historiography), we are still in the early days of developing methods to analyze how other groups, such as journalists and politicians, have incorporated the past into their narratives.

Identifying and interpreting the presence of the past in the discourse of non-historians is worthwhile, not because we expect to establish new facts about past events, but rather because laypersons and practitioners of other disciplines draw upon the past to make all kinds of judgments and decisions in daily life \cite{white2014practical}. The availability of large diachronic corpora has opened new avenues to study how and why people make reference to the past in their discourse (e.g. to convince others or to express emotion), and to analyze differences within particular time intervals as well as across time. The size of such corpora, however, combined with the scatteredness of references to the past, can make these corpora daunting to explore without the right tools.

In response to the ``spatial turn'' in Digital Humanities, substantial effort has been put into the development of tools that allow for spatial navigation through text collections. In the Pelagios project, for example, the Pleiades gazetteer serves to anchor locations mentioned in text to machine-readable representations of these locations, which can be combined with linked data to form rich map-based visualizations and allows for spatial access to the texts through the Peripleo search interface \cite{Rabinowitz2016}. The recognition that ``space and time are no more separate in human cognition
than they are in theoretical physics'' \cite[p. 43]{Rabinowitz2016} now motivates the development of tools that provide access to texts by the historical entities that they reference, to complement the evolving spatial approaches.

In this paper, we propose an approach to support researchers who aim to identify and interpret (indirect) references to historical periods in a particular discourse. It consists of a semantically-enhanced search engine that is able to find references to many entities at a time in diachronic corpora, which is combined with an interface that invites its user to actively sculpt the search result set.
As several pilot studies have indicated \cite{Olieman2017}, this approach is useful for those studying how groups of people pragmatically feature historical entities in discourse, remember or commemorate them, and understand their own identity in relation to these entities. Moreover, we are able to capture in linked data how definitions of historical periods are operationalized by researchers, and generate semantic annotations for the search results that users consider to be relevant.

\section{Searching for Colligatory Concepts}

The search for references to historical periods entails bridging the gap between conceptually-based information needs on one side, and term-based inverted indexes on the other. When a researcher is looking for the fragments of text that refer to, e.g., the French Revolution in a specific collection, they are not served well by providing only the documents that contain the literal phrase ``French Revolution.''
This is the case because such periods do not pre-exist in reality, waiting to be discovered and named, but rather come into being when the disparate observable elements of a phenomenon are ``seen together'' as a synthetic whole \cite{Golden2016}. Philosophers of science and of history have called this process colligation---a binding together of selected historical facts, culminating in the proposal of a ``colligatory concept'' which represents the historian's understanding of the facts as an `entwined whole' in a form that can be communicated to others \cite{Shaw2013}.

Periods are but one kind of colligatory concept that is commonly constructed by historians; the others being characters, such as `Louis XVI' and `the French people,' and ideal types, for instance `capitalism' and `revolution' \cite{Shaw2013}. Whereas characters are localized in time and space, and ideal types bind together certain aspects of the events in which these characters participate across time and space, periods are bound together by narratives that feature selected events which are distributed within (flexible) temporal and spatial bounds \cite{Shaw2013}. The PeriodO project has already produced a linked data gazetteer of periods as they are represented in the published works of historians and historiographers, using a nanopublication approach that connects the qualitative definition of a period by an individual author to its name, spatial extent, and a time interval, expressed together in RDF \cite{Rabinowitz2016,Golden2016}.

This model captures accurately the multivocal aspect of period definitions, but is not sufficient to retrieve references to these periods in discourse that is about another subject. Here we are confronted with the difference between colligatory concepts which are necessary constructs in the writing of history, and the \textit{subjects} that are intended to group together multiple histories that exhibit common ``patterns of colligation'' \cite[p. 1098]{Shaw2013}. As a colligatory concept, a period is a particular representation of the past that binds together historical entities in a unique narrative.
When this individual representation leads to further discourse, the discourse as a whole is not about the original colligation, but rather about a homonymous subject that allows us to, e.g, ask a librarian for `novels set during the French revolution.' In the practice of information organization, we establish common referents and shared structure between colligations in order to group a multiplicity of perspectives under a single label \cite{Shaw2013}.

We have no way of searching directly for the period that the researcher has in mind (it exists only as a cognitive representation), but by starting from its associated subject we can make an educated guess about the elements of the period.
Named entities---events, people, artifacts---are central in our search approach, because they serve as the points of consensus that enable the search for as of yet unknown perspectives on segments of the past. The aim is to collect (and represent) an intermittent discourse, consisting of sentences that make (indirect) reference to the target period of the search, with as much context as is needed to understand them.
To be sure, we do not equate the colligatory concepts that are put forward by historical work with the scattered references to the past that are found in non-historical discourse. Rather, we expect the (re)searcher to have a particular period (i.e. colligatory concept) in mind, and aim for the system to find references to the entities that are bound together by this period.

\section{Approach}

The task that we aim to facilitate---selecting a research corpus of text fragments that refer to a particular period---is not supported well by either subject indexing \cite{Shaw2013} or full-text search \cite{Hinze2016}, which nowadays provide the primary means of access to much of the text in archives and libraries. Subjects are traditionally assigned to whole documents. Our task, however, depends on the retrieval of individual sentences (and their context), given \textit{their subjects}. It seems infeasible to provide such granular access with manually assigned subjects, but recent advances in Entity Linking have enabled the automated identification of the referents in individual sentences. Even though entity linking systems are prone to errors that human annotators would not make, the entity links that they produce can still be useful to search for many entities simultaneously \cite{Olieman2017}. Semantically-enhanced search is made possible by incorporating entity links or similar semantic annotations into search indexes \cite{Hinze2016}. Our approach extends this practice by employing linked data to bridge the semantic gap between the search target (i.e. a period) and references to the entities that are bound together by this concept.

\subsection{Bootstrapping with DBpedia}
Some groupings of entities that people might want to search for already exist as Linked Open Data. The category network of Wikipedia serves an analog function to that of the subject access systems of librarians and archivists. It is a knowledge organization system (KOS) that relates more specific subjects to more general ones by \textit{broader-than} relations \cite{Shaw2016a}. As the product of the contributions of a diverse community, it encodes multiple perspectives on the world in a single structure which allows for multiple paths to exist between any two subjects.
We used DBpedia's RDF representation of this category network for our proof-of-concept, because we were working with a corpus that we enriched with entity links that point to DBpedia URIs. Its simple structure, represented in the SKOS ontology, may not be ideal for all research purposes, but it is sufficient to start a search process with a period-as-subject and invite the researcher to operationalize the particular period he/she wants to search for.


\begin{sloppypar}
In order to obtain possible mappings between periods of interest and the entities that are bound together by such periods, we extracted a subgraph of DBpedia, corresponding to Wikipedia's category network and its related entities, into a property graph database (see \cite{Rodriguez2015}). The category network is used at runtime to select potentially relevant entities given a root category, by traversing \code{skos:broader}\footnote{For namespace prefixes, see \url{https://dbpedia.org/sparql?nsdecl}.} and \code{dct:subject} relations in reverse direction. Starting, for example, from the root category \code{dbc:French\_Revolution}, the traversal would proceed through subcategories such as \code{dbc:Montagnards} and \code{dbc:French\_First\_Republic} to collect entities including \code{dbr:Reign\_of\_Terror}, \code{dbr:Maximilien\_Robespierre}, \code{dbr:Bastille}, and \code{dbr:Drownings\_at\_Nantes}.
\end{sloppypar}

Our proof-of-concept makes use of DBpedia, but any knowledge graph that conforms to the SKOS ontology can be loaded easily. Linked data that is structured differently can also be used, as long as a grouping or categorization that is familiar to the intended users can be derived from it. It is important that the representations of entities in this data are identifiable by the same URIs as those used in the entity links in the corpus, to be able to connect periods, via colligated entities, to the text fragments that refer to these entities.

Finally, the system needs access to coarse temporal clues about entities. Because DBpedia does not provide this data reliably across entity types, we extract mentioned years from the \code{rdfs:comment} values of DBpedia resources with a simple regular expression, and add them to the graph. The same technique may be successful for other linked data sources that provide textual descriptions which often include temporal expressions. It would be preferable, however, to use representations that incorporate structured temporal relations in the form of RDF literals for all entities.

\subsection{Search Interface Design}

Our proof-of-concept, named WideNet \cite{Olieman2017}, provides access to a semantically enriched version of the Dutch parliamentary proceedings---the ``verbatim'' records of the debates that take place in the Houses of Parliament (the \textit{Staten Generaal}).
These discussions touch almost on every issue that moved Dutch public opinion over more than two centuries.

\begin{figure}
  \includegraphics[width=.45\textwidth]{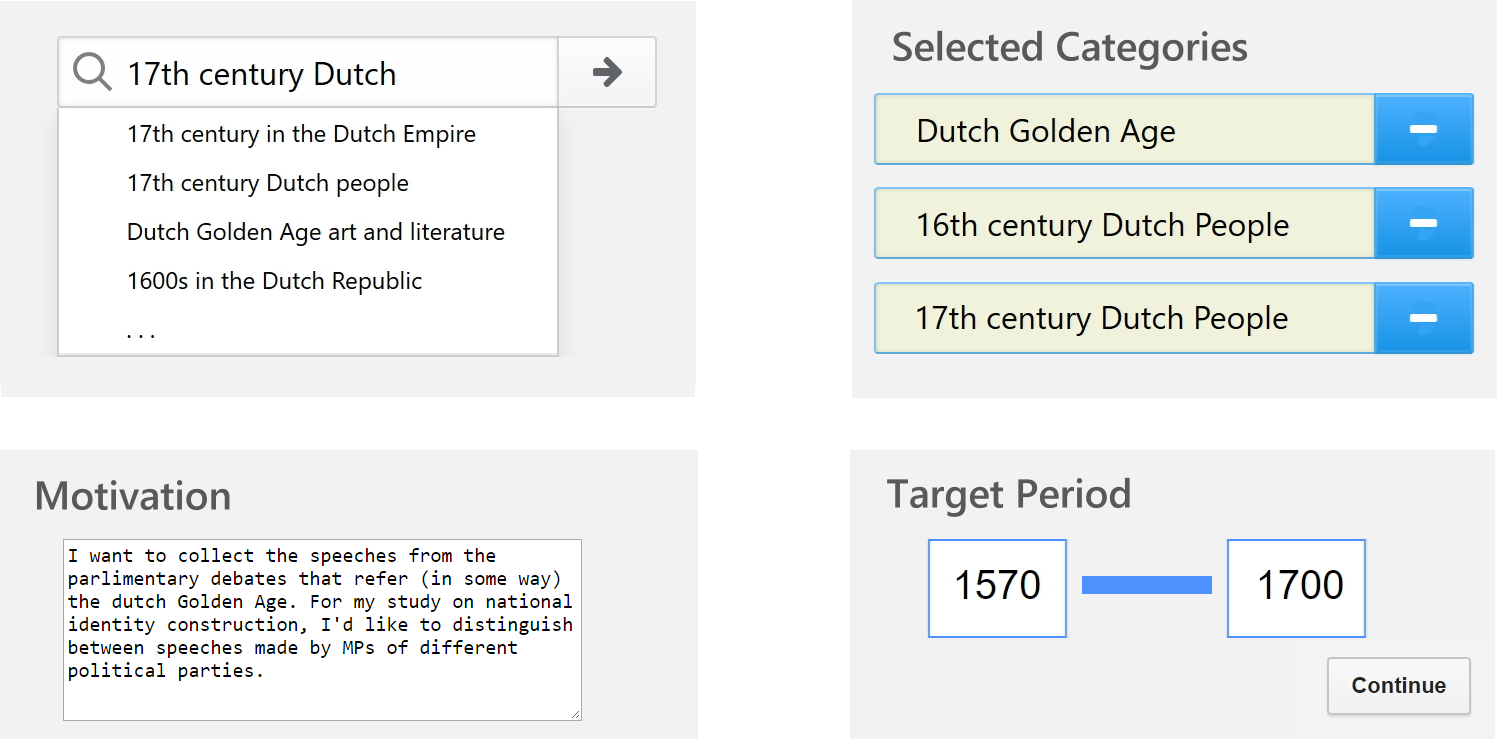}
  \caption{Initial query specification in WideNet.}
  \label{fig:ga-root}
\end{figure}

\begin{figure}
  \includegraphics[width=.45\textwidth]{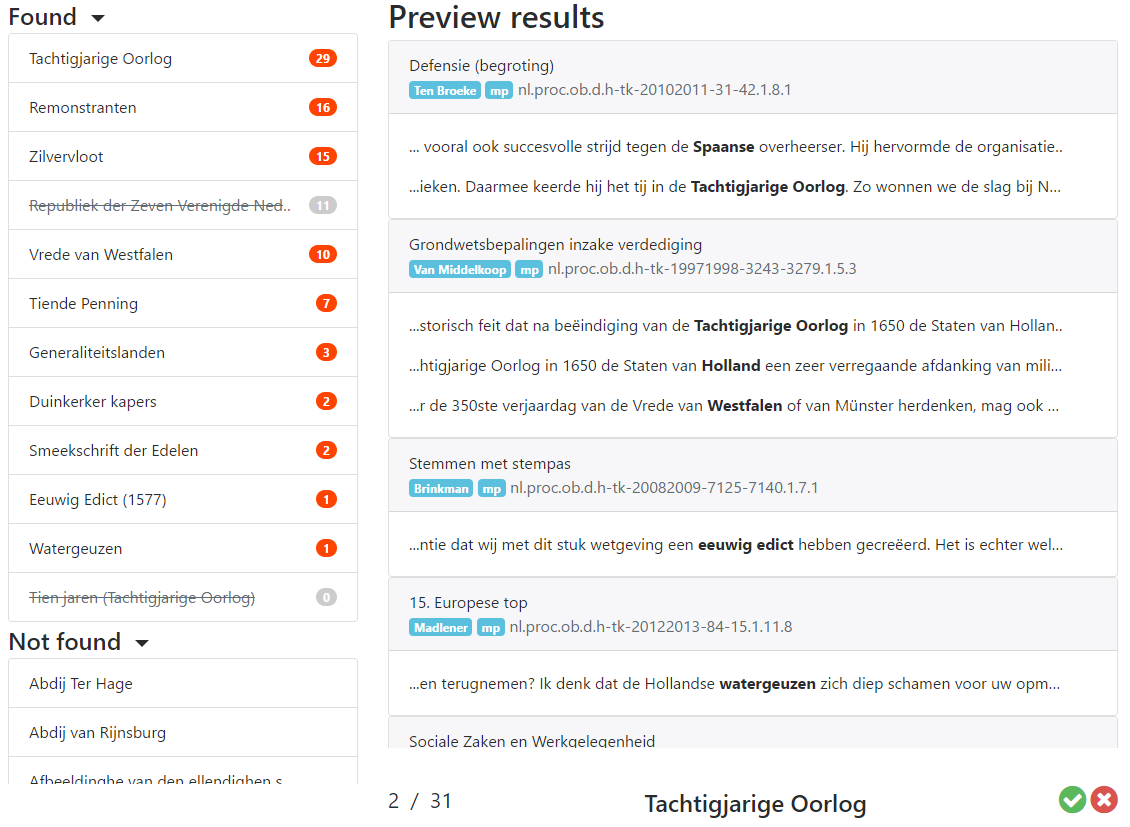}
  \caption{Assessing the relevance of categories and entities.}
  \label{fig:ga-preview}
\end{figure}

\begin{figure}
  \includegraphics[width=.45\textwidth]{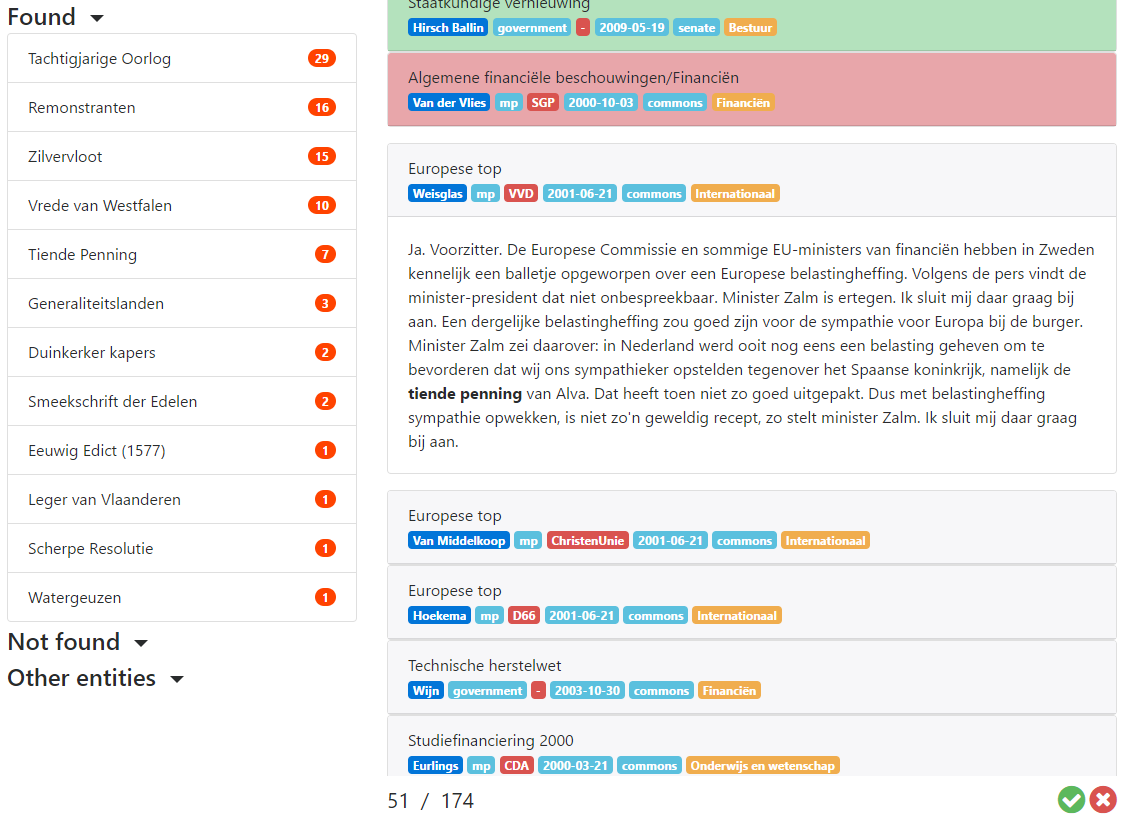}
  \caption{A closer look at the retrieved documents.}
  \label{fig:ga-reading}
\end{figure}

The interface guides users through three research phases: (1) selection of root category, (2) assessment of the categories' and entities' relevance, and (3) close-reading.
In the first step the user selects one or several root categories from a typeahead search box (see Figure \ref{fig:ga-root}), and demarcates the query by selecting a time period, which is used to prune the underlying entities of the selected categories.
WideNet subsequently retrieves the network of narrower categories for each selected root category, and collects the contained entities as potentially relevant query components. Behind the scenes, each entity is compared with the target period, and is considered to be outright relevant to the period, or not, or a borderline case, or as lacking temporal clues altogether. In the current implementation this classification is achieved with simple rules, based on the features: `fraction of years within period,' `fraction of intervals that overlap with the period,' and `has at least one year in period.'  The system uses this information to deselect (sub)categories where more than half of the dated member entities are out-of-period, i.e., those categories are excluded from the query.

The next step for the WideNet user is to assess which of the retrieved subcategories actually contain entities that lead to relevant results. By doing this, researchers can operationalize their own definition of their target period, at least for the purpose of retrieval. The interface facilitates this task by showing, per subcategory, which entities are mentioned in the corpus, and how frequently, as well as which entities did not occur (see Figure \ref{fig:ga-preview}). It also displays a list of preview results, showing limited context, to offer quick clues about the relevance of the category. This preview is also useful to identify individual entities that are not relevant after all, which can be deselected by the user. At the end of this step, the researcher's decisions amount to a motivated and organized representation of `relevant' or `possibly present' elements of the target period.

After inspecting and selecting relevant categories of entities---thereby sculpting the final query, the WideNet interface allows further scrutinizing of the sources by providing an environment in which the retrieved documents can be studied up-close, as shown in Figure \ref{fig:ga-reading}. By situating the close reading activity within the same interface, the user is able to compile a corpus of relevant documents which may be saved and exported.
Moreover, the user can examine the results in relation to the document metadata, e.g. to look for saliency by plotting the annotations over time, or to study bias by comparing how often different political parties refer to the entities of interest. The selected corpus, representing a fragmented discourse, may also be used to analyze changes in how and why the demarcated segment of the past was evoked, establishing contesting perspectives and trends over time, provided that enough references could be found.

\shrink
\subsection{Capturing Reusable Data}

As a product of the search, semantic annotations are created which link the source documents to the entities that are referenced in the corpus. These expert-approved annotations have much more value than the automatically generated entity links. For one, the identity of the referents is established with a greater confidence when a user chooses to include a particular document fragment in his/her research corpus. We cannot interpret this as a direct assessment of the entity links, but when a user has confirmed that the document fragment (indirectly) refers to the target period, it is tempting to assume that the document was retrieved because the entity links are correct, and not by some mistake. To model the provenance of such relevance decisions, it is necessary to produce a representation of the broader context of the search process to which individual assertions can be explicitly related (see \cite{Golden2016}).

In the first screen that users encounter when starting a new search process in WideNet (depicted in Figure \ref{fig:ga-root}), we are able to capture the motivation and a rough demarcation of the search. This information forms the backbone of a representation to which subsequent assertions can refer. The selection of categories and entities (in Figure \ref{fig:ga-preview}) provides assertions about their relevance for the search target, according to the researcher and dependent on the corpus. We currently store each decision that is made by the researcher, so when a previously made decision is revisited, we create representations of e.g. both the act of deselecting an entity, and reselecting it after more preview results had been inspected. Capturing this process data, rather than only the final selection of entities, benefits the richness of the provenance of the subsequent assertions about the relevance of text fragments.

The semantic annotations that are derived from relevance assertions on text fragments can provide richly indexed access to statements about the past. This notion is similar to Ryan Shaw's proposal for ``deep gazetteers,'' in which multiple descriptions of the same named entity are linked to fragments of discourse in which its name is used \cite{Shaw2016a}. In our case, however, we use representations of periods-as-subjects to connect particular conceptions of these periods to discourse that refers to these periods from a different perspective, rather than use of the same name per se. Our approach can produce ``projections'' of the parts of the past that were present in a particular discourse, which can be organized according to the KOS that was used to perform the search, but also according to any other KOS that includes representations of the same entities, provided that equivalence of identity (e.g. \code{owl:sameAs}) can be established between the knowledge organization systems.

While we capture the described representations of search processes and semantic annotations in our current proof-of-concept, we do not yet publish this data. Our approach is related to ongoing efforts to produce reusable data for research in the humanities (e.g. \cite{Shaw2015,Golden2016,Shaw2016a}) and we are still investigating how we can best link to the existing models. We have prioritized the development of a useful search tool over the production of reusable data in order to investigate which data can be captured during actual research in the humanities, rather than designing our models first and finding out later that they are not as usable as we had hoped \cite{Shaw2015}.

\section{Conclusion and Outlook}

In searching diachronic corpora for fragmented discourses about historical periods, we need to come to a fundamental understanding of conceptual difference before we can give any account of conceptual drift over time. Our approach for finding references to periods consists of a semantically-enhanced search engine that is able to find references to many entities at a time in diachronic corpora, which is combined with an interface that invites its user to actively sculpt the search result set. 
Besides yielding sources that are useful for the searcher, the search tool also produces an operationalization of the search target by the (re)searcher as well as semantic annotations that are much richer than those that we can generate algorithmically.

Although a pragmatic treatment of concepts is sufficient to search for multifaceted subjects, we envision how the products of such search processes can collectively provide a shared source for more elaborate knowledge representations.
In designing a tool that implements this approach, we were faced with some trade-offs between usability of the tool and reusability of the data it captures. We prioritize supporting the present-day researcher well, and facilitate publishing the search process and its results as linked open data, so that subsequent refinement of the captured data may be a community effort.

\begin{acks}
We thank the anonymous reviewers for their suggestions and remarks.
This work was supported by the Netherlands Organization for Scientific Research (ExPoSe project, NWO CI \# 314.99.108).
\end{acks}

\bibliographystyle{ACM-Reference-Format}
\bibliography{sigproc} 

\end{document}